\documentclass[onecolumn]{aa} 
\usepackage{graphicx}
\usepackage{subfigure}
\usepackage[]{natbib}
\usepackage{threeparttable}

\begin{document}
   \title{The structure of TeV-bright shell-type supernova remnants}


   \author{
    Chuyuan Yang\inst{1}\inst{2}
    \and
     Siming Liu\inst{3}
     \and
     Jun Fang\inst{4}
     \and
     Hui Li\inst{5}
     }
   \institute{ Yunnan Observatories, Chinese Academy of Sciences, Kunming 650011, China, chyy@ynao.ac.cn
   \and
    Key Laboratory for the Structure and Evolution of Celestial Objects, Chinese Academy of
     Sciences, Kunming 650011
     \and
     Key Laboratory of Dark Matter and Space Astronomy, Purple Mountain Observatory, Chinese Academy of Science, Nanjing 210008, China; liusm@pmo.ac.cn
     \and
     Department of Astronomy, Yunnan University, Kunming 650091, China, fangjun@ynu.edu.cn
   \and
   Los Alamos National Laboratory, Los Alamos, New Mexico, 87545, USA
   }
\titlerunning{Structure of TeV-bright supernova remnants}

\date{Received / Accepted}


  \abstract
{}
{Two-dimensional magnetohydrodynamic (MHD) simulations are used to model the emission properties of TeV-bright shell-type supernova remnants (SNRs) and to explore their nature.}
{ In the leptonic scenario for the TeV emission, the $\gamma$-ray emission is produced via Inverse Compton scattering of background soft photons by high-energy electrons accelerated by the shocks of the SNRs. Since the energy density of the Cosmic Microwave Background (CMB) radiation and that of the IR/Optical background photons are much higher than that of the photons produced by the same high-energy electrons via the synchrotron process, in a previous paper, we showed that the observed correlation between X-ray and TeV brightness of SNR RX J1713.7-3946 can be readily explained with the assumption that the energy density of relativistic electrons is proportional to that of the magnetic field. The TeV emissivity is therefore proportional to the magnetic field energy density and MHD simulations can be used to model the TeV structure of such remnants directly. 2D MHD simulations for SNRs are then performed under the assumption that the ambient interstellar medium is turbulent with the magnetic field and density fluctuations following a Kolmogorov-like power-law spectrum.
}
{(1) As expected, these simulations confirm early 1D and 2D modelings of these sources, namely the hydrodynamical evolution of the shock waves and amplification of magnetic field by Rayleigh-Taylor convective flows and by shocks propagating in a turbulent medium;
(2) We reproduce rather complex morphological structure for $\gamma$-rays, for example, bright thin rim and significant asymmetry, suggesting intrinsic variations of the source morphology not related to the structure of the progenitor and environment;
(3)
  The observed radial profile of several remnants are well reproduced with an ambient medium density of $0.1-1$ cm$^{-3}$. An even lower ambient density leads to a sharper drop of the TeV brightness with radius than what is observed  near the outer edge of these remnants.
 }
{In a turbulent background medium, we can reproduce the observed characteristics of several shell-type TeV SNRs with reasonable parameters except for a higher ambient density than that inferred from X-ray observations.
}
   {}


   \maketitle
%

\section{Introduction}
It is generally believed that supernova remnants (SNRs) are
the main acceleration sites of Galactic cosmic rays. Since charged cosmic ray particles escaping from SNRs are mixed in the Galactic
magnetic fields and can not be used to image their sources of origin directly,
detection of high-energy $\gamma$-rays from SNRs is very
useful for testing whether SNRs are responsible for the origin
of the bulk of the cosmic rays observed at the Earth.
However, an open question is whether the observed high-energy
$\gamma$-ray emission from an SNR has a hadronic or a leptonic
origin \citep[e.g.,][]{kw08, ylb12}.

Nonthermal photons from SNRs in the radio to X-ray
bands are produced by the synchrotron radiation of relativistic
electrons. The detection of synchrotron X-rays in some
SNRs shows that electrons can be accelerated efficiently to TeV energies by shocks of SNRs \citep[e.g.][]{kyet95}.  Although several young
SNRs have been detected to emit GeV to TeV  $\gamma$-rays \citep[e.g.][]{ab11,ak11}, it is not clear whether these SNRs produce
high-energy protons or not.
High-energy $\gamma$-ray photons can be produced
by either inverse-Compton(IC) process or $\pi^0$-decay
corresponding to the leptonic or hadronic origin, respectively.
For example, $\gamma$-ray emission from SNR RX J1713.7-3946 is likely produced by very energetic electrons radiating through IC scattering against some
photon background \citep{ab11}, and objects like the Tycho SNR likely emit GeV $\gamma$-rays via the decay of neutral pions produced in nuclear collisions between shock accelerated relativistic nuclei and those in the background plasma \citep{mc12, ac13}.

Recent X-ray and $\gamma$-TeV observations have revealed a group of TeV-bright shell-type SNRs with SNR RX J1713.7-3946 as a prototype. Similar to SNR RX J1713.7-3946, no thermal emission has been detected from SNRs RX J0852.0-4622 and HESS J1731-347; the $\gamma$-ray luminosity is only a factor of a few lower than the X-ray luminosity; and they have a very hard spectrum in the $\sim$ GeV energy range, which favor the leptonic origin for the GeV$-$TeV emission with a low mean magnetic field of $\sim 10\ \mu$Gauss \citep{ll11, e12, ylb12, yz14}.
Rapid variations of X-ray filaments with a width of about 0.1 light year on a timescale of $\sim 1$ year have been interpreted as due to radiative energy loss of TeV electrons in strong ($\sim 1$ mG) fields \citep{u07}. The strong field may only exist in these filaments with a very small volume filling factor, and alternative explanations for this variability with a weak mean field have been proposed as well \citep{bu08, lf08}. Although thermal emission has been detected from SNR RCW 86 \citep{w11}, the characteristics of nonthermal emission from this remnant is very similar to other TeV-bright shell-type SNRs \citep{l12, yh14}. These results imply a universal mechanism of particle acceleration by shocks of SNRs \citep{ylb12}.

The nonthermal emission of these remnants has a well-defined shell structure and there is evidence that the relative mean thickness of the emitting shell varies from remnant to remnant \citep{ah06, ah07}. Previously 1D modeling of particle acceleration in SNRs has been used to study the radial brightness profile in the hadronic scenario \citep{bv10}. Detailed study of the source structure with MHD simulations hasn't been carried out.
Although these remnants have complicated appearance in X-rays and $\gamma$-rays with significantly fluctuations in both the radial and azimuthal directions, there is a good correlation between the X-ray and $\gamma$-ray brightness \citep{ac09}.
Magnetic fields play an important role in collisionless astrophysical shocks and in the acceleration of charged particles. They can be amplified by cosmic ray induced streaming instability in the presence of efficient particle acceleration \citep[e.g.][]{b04,rs09}. In a turbulent medium, they can also be amplified by turbulent motion in the shock downstream\citep{gj07}. The latter has been explored recently by \citet{gs12} who performed extensive 2D numerical simulations for SNR blast wave interacting with a turbulent plasma background and found that the magnetic field can also be amplified by Rayleigh-Taylor convective flows induced at the contact discontinuity of the shock flows.
In our previous paper \citep{yl13}, we have shown that the observed correlation between X-ray and $\gamma$-ray brightness for SNR RXJ 1713.7-3946 suggests that the energy density of the accelerated electrons is proportional to that of the magnetic field in the leptonic scenario for the $\gamma$-ray emission. In this paper, we study the structure of these remnants and the effect of magnetic inhomogeneity on the TeV profile and morphology  by using 2D MHD simulations.

Our 2D MHD simulations are based on the PLUTO code \citep{m07}.
In Section 2, we describe the MHD model and show the magnetic field structure.
In section 3, an algorithm is proposed to convert the 2D magnetic field structure into a $\gamma$-ray image of the remnant and several simulations are done to fit the azimuthal averaged radial profile of the TeV brightness of the remnants mentioned above. Discussion and our conclusions are given in Section 4.


\section{ Model Description}
\label{models}

The numerical model has been described in \citet{yl13}. For the sake of completeness, we briefly summarize the key issues here.
The dynamical evolution of an SNR shock propagating into a turbulent ambient medium is simulated with
the time-dependent ideal MHD equations of mass, momentum, and energy conservation:
\begin{equation}
\partial_t \rho+\nabla\cdot(\rho\bf u ) =0\,,
\end{equation}
\begin{equation}
\partial_t \rho{\bf u}+\nabla\cdot\left(\rho{\bf uu-{BB\over 4\pi}}\right)+ \nabla P^\prime =0 \,,
\end{equation}
\begin{equation}
\partial E +\nabla\cdot\left[(E+P^\prime){\bf u-{B(u\cdot B)\over 4\pi}}\right]=0\,,
\end{equation}
and the induction equation:
\begin{equation}
\partial_t {\bf B}+\nabla\times({\bf u\times B }) =0 \,,
\end{equation}
where $P^\prime=P+B^2/8\pi$ is the total pressure with thermal pressure $P$ and magnetic pressure $B^2/8\pi$, and cgs units have been adopted in this paper.
$E$ is the total energy density:
\begin{equation}
E=\frac{P}{\gamma-1}+\frac{1}{2}\rho u^2+\frac{B^2}{8\pi},
\end{equation}
$\rho$, $\bf u$, $\bf B$, and $\gamma=5/3$ are the plasma mass density, fluid velocity, magnetic field, and the adiabatic index, respectively.



The numerical scheme to solve the ideal MHD equations was presented by \citet{m07} in detail.
We model the simulation in a two-dimensional Cartensian coordinate ($x, y$) with uniform grids.  The size of simulation domain is chosen to be 40 pc $\times$ 40 pc to cover the extension of young SNRs . The supernova blast wave is driven by the injection of internal energy and ram pressure in a small circular region at the center of simulation box, in which the density is assumed to be constant corresponding to a plateau volume ($4\pi r_{\rm inj}^3/3$) with $r_{\rm inj}=0.5$ pc. The initial magnetic field and density in the background plasma include an average component and a turbulent component. We assume a constant average magnetic field $B_0$  along the $x$ direction and a constant average gas density $n_0$ \citep{gs12}.

Both density and magnetic fluctuations are generated by the assumption of a Kolmogorov-like power-law spectrum of the form
\begin{equation}
P(k) \propto \frac{1}{1+(kL)^\Gamma},
\end{equation}
where the spectral index $\Gamma$ depends on the dimensionality and equals $8/3$ for 2D system.
$k$ is the magnitude of the wavevector and $L=3\ \mathrm{pc}$ is the turbulence coherence length. The turbulence
is generated by summing a large number of discrete wave modes with random phases \citep{gj99}.
The random component of magnetic field is given by
 \begin{eqnarray}
\delta \textbf{B} (x, y) = \sum^{N_m}_{n=1} &\sqrt{C_B2\pi k_n \Delta k_n
P_B(k_n)} (\sin \theta_n \hat{x} - \cos \theta_n \hat{y})  \nonumber \\
& \times\exp (i \cos \theta_n k_n x + i \sin \theta_n k_n y + i \phi_n)
\end{eqnarray}
where $P_B(k_n)$ represents the power of the wave mode $n$ with wavenumber $k_n$. The turbulence distributes randomly in propagation direction $-1<\cos\theta_n<1$ and with a random phase $0<\phi_n<2\pi$. $C_B$ is a normalization constant determined by $\langle \delta B^2\rangle=B_0^2$, where $\langle\rangle$ represents averaging in space.

The density fluctuations satisfy the following probability distribution \citep{gj07,bl00}:
\begin{equation}
n(x,y)=n_0\exp(f_0+\delta f)
\end{equation}
where $f_0$ is a constant chosen to give the average density $n_0$ and the description of $\delta f$ is similar to the turbulent part of the magnetic field:
\begin{eqnarray}
\delta f(x, y) =&\sum^{N_m}_{n=1} \sqrt{C_f2\pi k_n \Delta k_n P_f(k_n)}  \nonumber \\
&\times\exp(i \cos \theta_n k_n x + i \sin \theta_n k_n y + i \phi_n)\,,
\end{eqnarray}
with the normalization constant $C_f$ determined by $\langle(\delta n)^2\rangle=0.4n_0^2$.


\citet{gs12} carried out extensive study of 2D MHD simulations of SNRs in a turbulent medium. We first have a test run with $n_0=0.04$ cm$^{-3}$, $B_0=0.8\ \mu$G, the total injected internal energy $E_{SN}=2\times10^{51}$ ergs, and the total mass of the ejector $M_{ej}=2M_\odot$. The left panel of Figure \ref{test} shows the magnetic field structure at 1600 years. The units of the coordinates are pc and the color scale is the logarithmic of the magnetic field in units of $\mu$G with the peak magnetic field reaching about 100 $\mu$G.
  The magnetic field near the shock front is mostly amplified by turbulent motion in the downstream \citep{gj07}. The strongest magnetic field however is produced by Rayleigh-Taylor convective flows near the contact discontinuity. Since the ejector is uniform, the magnetic field is very weak in the downstream of the reverse shock.

 In order to construct the profile of $\sim$TeV emission, we adopt the leptonic scenario where the $\gamma$-ray is produced via the IC of background soft photons off high-energy electrons accelerated by the shocks. Besides the CMB radiation, we also include an IR background photon component with a temperature of $T=30$ K and an energy density of 1.2 eV cm$^{-3}$ \citep{yl13}. The energy density of radio to X-ray band synchrotron component of the SNRs is much lower than the energy density of the CMB radiation, the synchrotron self-Comptonization process therefore can be ignored. Since the background photon field is approximately uniform, the $\gamma$-ray emission characteristics are then determined by the properties of energetic electrons. \citet{yl13} showed that the observed correlation between the X-ray and $\gamma$-ray brightness of these SNRs implies that the electron energy density is proportional to the magnetic field energy density, we then can describe the electron spatial distribution simply as the following:
\begin{equation}
\frac{dN(x,y)}{d\gamma_e}=CB(x,y)^2 F_{\gamma_e} ,
\label{ng}
\end{equation}
where $F_{\gamma_e}\sim \gamma_e^\delta \exp(-\gamma_e/\gamma_{\rm max})$ is a power law energy distribution with an exponential cutoff and $\gamma_e$ is the electron Lorentz factor. From the overall $\gamma$-ray spectrum of SNR RX J1713.7-3946, we find $\delta=-2$ and $\gamma_{\rm max} = 1.3\times 10^7$ \citep{ll11, yl13}.  The proportionality of the magnetic field and high-energy electron energy densities then implies that the electron distribution only depends on the magnetic field energy density and the structure of TeV emission is determined by $B^2$.  A recent study of TeV bright shell-type SNRs by \citet{yz14} showed that the energy partition between energetic electrons above 1 GeV and the magnetic field may be different for different SNRs with a tendency that the total magnetic field energy within the volume enclosed by the front shock of SNRs increases faster than that of energetic electrons as the SNRs evolve. But the proportionality of these energy densities is inferred from the correlation between the X-ray and $\gamma$-ray brightness in the leptonic scenario, we therefore expect our treatment be valid for sources with a good correlation between the X-ray and $\gamma$-ray brightness as far as the spatial structure of the $\gamma$-ray emission is concerned.

\section{Results}
In the following, we use SNR RX J1713.7-3946 as a prototype to describe the modeling. The remnant was first discovered in X-rays with the
ROSAT all-sky survey in 1996 \citep{pa96}. It is close to the Galactic plane and its distance and ages are about 1 kpc and 1600 yrs respectively \citep[e.g.][]{f03,cc04}, and the radius of the remnant is about 10 pc \citep{ac09}. The gamma-ray image of SNR RX J1713.7-3946 is consistent with a thick emitting shell with a thickness close to half of the radius and with the emissivity of southeast part is a factor of two lower than the rest \citep{ah06}.

\subsection{Structure and Radial Profile of TeV Emission}

\begin{figure}[htb]
\centering
\includegraphics[width=4 cm]{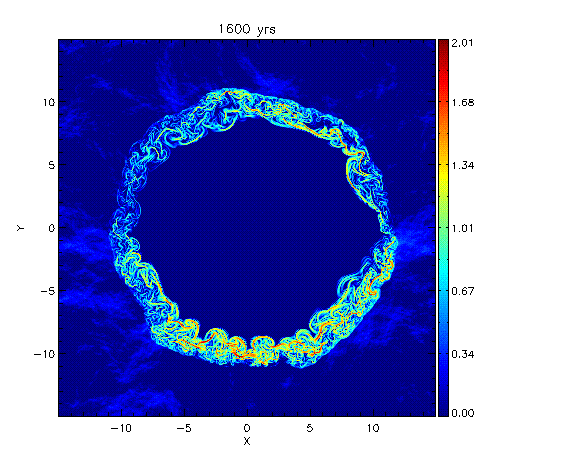}
\includegraphics[width=5 cm]{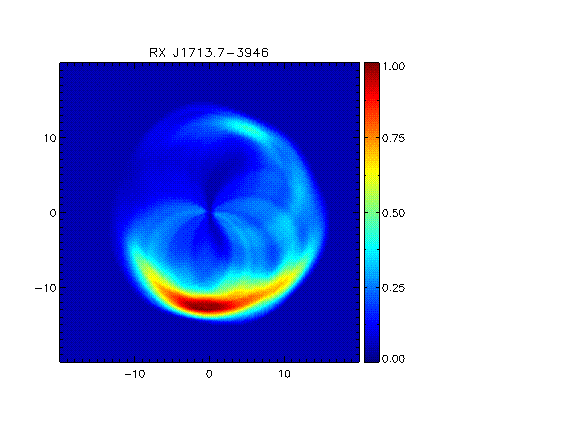}
\includegraphics[width=5 cm]{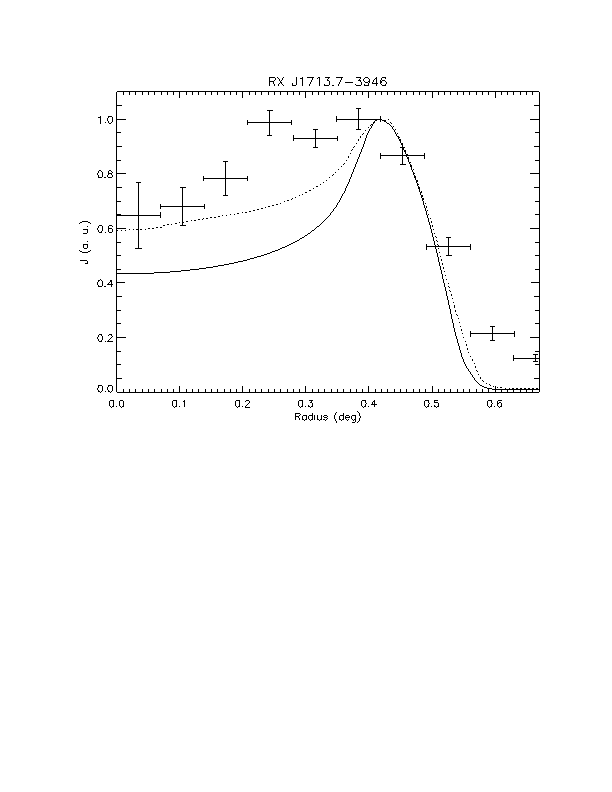}
\caption{Left: The structure of the magnetic field $B$ in units of $\mu$G at 1600 years of a test run with $n_0=0.04$ cm$^{-3}$. The rest of the model parameters are shown in the text. Middle: TeV image derived from the 2D MHD simulation. The image has been normalized to its peak value. Right: Comparison of the observed radial profile averaged along the azimuthal direction with results (solid line) derived from the 2D MHD simulation.  The dotted line is obtained by averaging the radial brightness profiles obtained in the position angle range of $0.7\pi$ to $0.95\pi$ in the middle panel.
\label{test}}
\end{figure}
To model the observed morphology of the TeV emission, we also need to construct a pseudo 3D structure with 2D simulations.
One can integrate $B^2$, which is proportional to the TeV emissivity, along a direction (presumably corresponding to the line-of-sight) in the 2D simulation domain to obtain a brightness profile in the radial direction of the SNR. Such a radial profile can be obtained for all angular directions in the 2D domain. To mimic the observed TeV image, we stack the radial profile obtained above continuously in a polar coordinate and smooth the resultant 2D structure with the point spread function of the HESS. The middle panel of Figure \ref{test} shows such an image derived from the structure of the magnetic field shown in the left panel for an assumed distance of 1 kpc. The arcs connecting the center and the outer rim are artifacts of the algorithm proposed above. Although this image does not recover all details of the TeV image obtained with the HESS \citep{ah06}, the strong variation of the brightness in the azimuthal directions and the shell structure is well reproduced.  The one-sided structure of the simulated TeV image is closely related to the asymmetry of the magnetic field structure. Due to the presence of a large scale magnetic field in the horizontal direction, the magnetic field at the bottom and top segments of the SNR shown in the left panel is preferentially amplified with the bottom segment amplified the most. Since the TeV emissivity is proportional to $B^2$, the asymmetry in the magnetic field structure is further amplified in the TeV image giving rise to the bright southern rim in the TeV image shown in the middle panel.

The right panel of Figure \ref{test} compares the model predicted azimuthal averaged radial profile with the observed profile. The radial profiles have been normalized to their peak value. It can be seen that the solid line of the model result is much narrower than the observed profile. This is not surprising since the left panel of Figure \ref{test} shows that the emission comes from a shell with a thickness less than one-third of the radius while the observed profile is consistent with an emitting shell with a thickness about half of the radius \citep{ah06}.
In the 1D model proposed by \citet{bv10}, it is claimed that a broad radial profile can be produced by considering the angular resolution of the observations. This 1D model predicts a very bright rim at the shock front. The smoothing effect caused by the angular resolution of the observations can reduce the brightness of this rim dramatically and therefore suppress the brightness contrast between the center and the shock front regions. In our 2D model, we don't have a very bright rim and the azimuthal variance also reduces the intrinsic brightness contrast of the averaged radial profile. The angular resolution of the observations does not change the brightness contrast between the center and the shock front regions drastically.

Considering the complex structure of the magnetic field in the left panel Figure \ref{test} and its dependence on the large scale magnetic field, projection effect may play a role in the brightness profile. Instead of averaging of the whole $2\pi$ angle, the dotted line shows the profile averaged from $0.7\pi$ to $0.95\pi$. The model prediction is more consistent with observations than the solid line. However, the brightness profile of the model still falls faster than the observed profile near the edge of the remnant. 3D simulations are needed to clarify this issue.

\subsection{Application to TeV Images of SNRs}



\begin{table}[htb]
\begin{center}
\caption{Model Parameters.\label{tbl-input}}
\begin{threeparttable}
\begin{tabular}{ccccccccccc}
\hline\hline Cases &$E_{\rm SN}$ & $M_{\rm ej}$ &$R_s$ & $n_0$  & Age &$V_s$ & $D$& Ref.& \\
                  & (erg) & ($M_{\sun}$) & (pc) & ($\rm cm^{-3}$) &(year) &  (km/s) & (kpc) & &\\
\hline
RX J0852.0-4622 & $1\times 10^{51}$ & 1.4 & 8 & 0.1& 990 & 5200& 0.5& 1-3& \\
 RX J1713.7-3946 &  $2\times 10^{51}$ & 2 & 9 & 0.8 &  1600 &  3700& 1.0 & 4-7  &\\
RCW 86 & $1\times 10^{51}$ & 1.4 & 11 & 0.4 & 1800 &  3900& 1.9& 8-11 &\\
  HESS J1731-347 & $1\times 10^{51}$ & 2 & 13 & 0.1& 2000 &  4300& 3.0&12-15 &\\
\hline
\end{tabular}
\begin{tablenotes}
  \item[] Reference - (1) \citet{ah05}; (2) \citet{ah07}; (3) \citet{byh05}; (4) \citet{f03}; (5) \citet{cc04}; (6) \citet{ac09}; (7) \citet{ah06};
  (8) \citet{ah09}; (9) \citet{r96}; (10) \citet{bet00}; (11) \citet{v06}; (12) \citet{tian08}; (13) \citet{ah08};  (14) \citet{ak11}; (15) \citet{bm12}
\end{tablenotes}
\end{threeparttable}
\end{center}
\end{table}




\begin{figure}[htb]
\centering
\begin{minipage}{0.2\textwidth}
\includegraphics[width=1.2\textwidth]{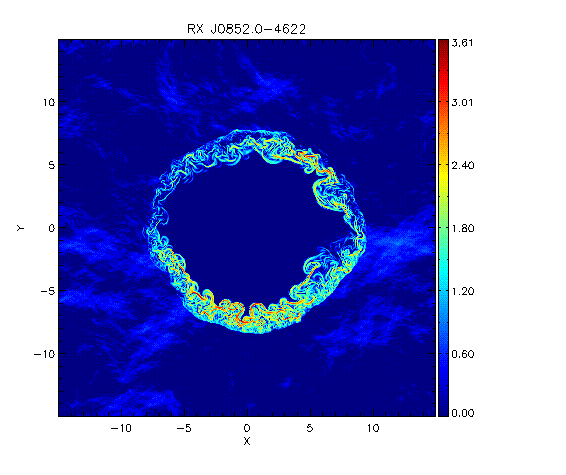} \\
\includegraphics[width=1.2\textwidth]{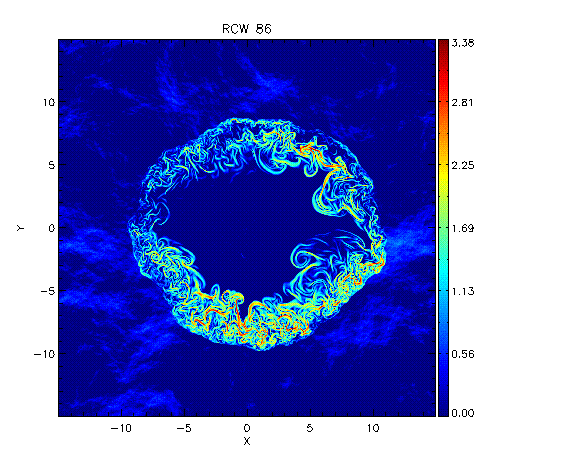}
\end{minipage}
\begin{minipage}{0.2\textwidth}
\includegraphics[width=1.2\textwidth]{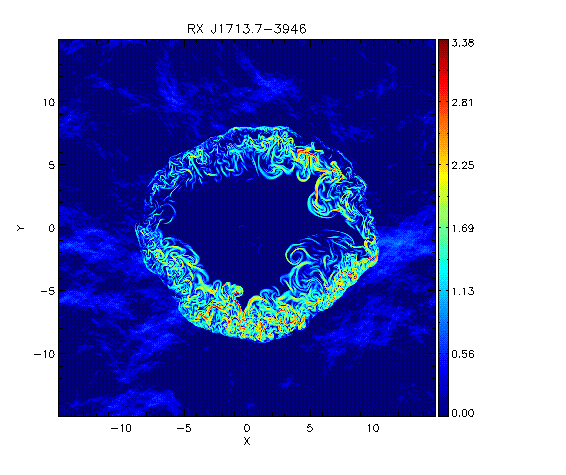} \\
\includegraphics[width=1.2\textwidth]{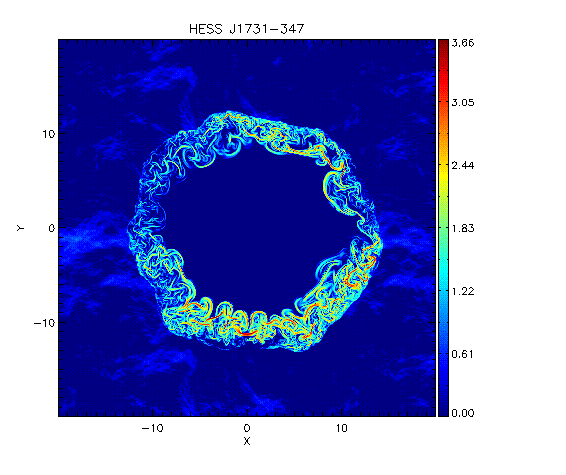}
\end{minipage}
\caption{The structure of magnetic energy density $B^2$ for four SNRs investigated in this paper. The parameters of model are described in Table \ref{tbl-input}.}
\label{mgn}
\end{figure}

\begin{figure}[htb]
\centering
\begin{minipage}{0.2\textwidth}
\includegraphics[width=1.5\textwidth]{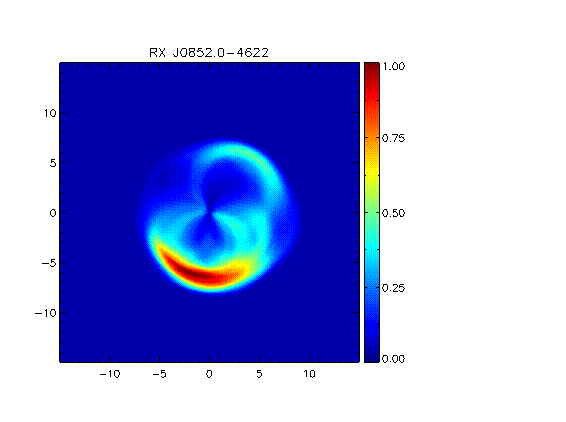} \\
\includegraphics[width=1.5\textwidth]{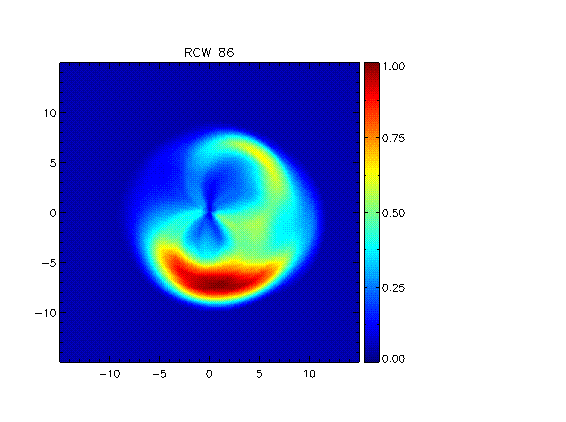}
\end{minipage}
\begin{minipage}{0.2\textwidth}
\includegraphics[width=1.5\textwidth]{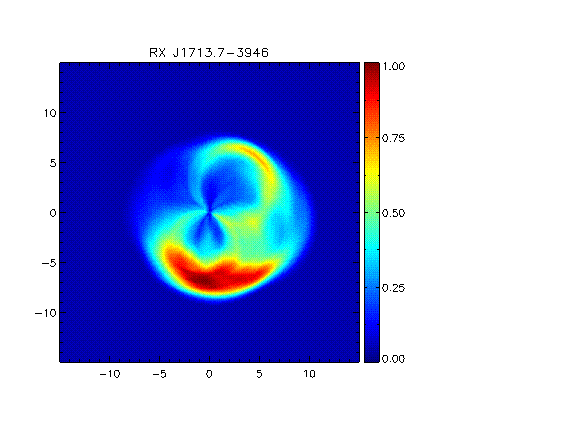} \\
\includegraphics[width=1.5\textwidth]{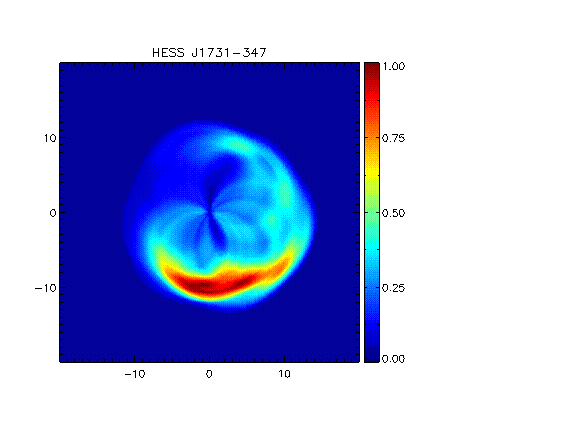}
\end{minipage}
\caption{Same as the middle panel of Figure \ref{test} but for the four SNRs in Figure \ref{mgn}.
}
\label{mpy}
\end{figure}

\begin{figure}[htb]
\centering
\includegraphics[width=8 cm]{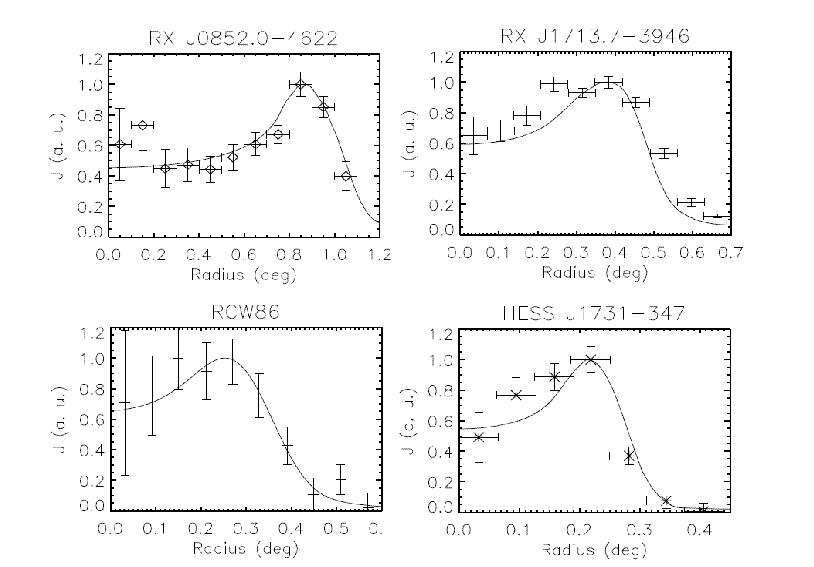}
\caption{Same as the right panel of Figure \ref{test} bur for the four SNRs in Figure \ref{mgn}.
The data of RX J0852.0-4622, RX J1713.7-3946, RCW86, and HESS J1731-347 are from HESS observations \citep{ah06,ah07,ah09,ak11}.
\label{pdf}}
\end{figure}

Besides SNR RX J1713.7-3946, SNR RX J0852.0-4622 is another very important Galactic SNR. Its X-ray and gamma-ray images are consistent with a very thin shell of $1^{\circ}$ in radius and $\sim 0.2^\circ$ in thickness.
 Its radius is $5- 10$ pc\citep{ah05,ah07} for a distance of $0.26-0.50$ kpc, and its age is about $420-1400$ yrs \citep{byh05}. The young shell-type SNR RCW 86 (also known as G315.4-2.3 and MSH 14-63) has been observed in radio
 \citep{kc87}, optical \citep{s97}, X-ray
\citep{p84}, and gamma-ray \citep{yh14, ah09} with a nearly circular shape of $40^\prime$ in diameter.
\citet{r96} deduced a distance of
2.8 kpc and an age of $\sim 10, 000$ years by using optical observations, whereas recent observations of the northeast part of the remnant with the Chandra and XMM-Newton satellites strengthen the case that the event recorded by the Chinese in 185 AD was a supernova and RCW 86 is likely its remnant \citep{v06}.
\citet{tian08} reported the discovery of a faint radio and X-ray shell with a distance of $3.2\pm0.8$ kpc closely matching the extended TeV source HESS J1731-347 \citep{ah08}.
Deeper $\gamma$-ray observations of the source have revealed a large shell-type structure with similar position and extension ($r\sim 0.25^{\circ}$) as the radio SNR, confirming their association \citep{ak11}. These authors suggested that its distance is not much larger than the derived low limit of $3.2$ kpc, and a physical size of $\leq 15$ pc is also reasonable.

 To improve the fitting to the azimuthal averaged radial profile of the TeV brightness, one may adjust the mean density $n_0$ of the ambient medium.
In the following, we will adjust $n_0$ as the major free parameter to fit the radial brightness profile of these SNRs.
Considering the ambiguity of the progenitor, we picked the mass and total energy of the ejector in a reasonable region to reproduce the physical size of these SNRs. A fit to the observed radial TeV brightness profile is then used to constrain the density $n_0$. The model parameters are shown in Tab.\ref{tbl-input} including the output parameters, such as the shock velocity $V_s$ and age. 

Figure \ref{mgn} shows the magnetic energy density. It is evident that the shell thickness increases with the increase of the mean density of the background.  The results for RX J1713.7-3946 and RCW 86 have thicker shells than the other two sources. Figures \ref{mpy} and \ref{pdf} show that the TeV morphology and radial profile of these sources can be approximately reproduced with our model. To improve this model, one may consider more realistic 3D simulations with the structure of the progenitor wind and environment of these sources taken into account properly. We note that the background density derived from our model is higher than the upper limit inferred from the X-ray observations, which poses the most serious challenge to our model. The thermal X-ray emission must be suppressed in the shock downstream to be consistent with the absence or weakness of thermal X-ray emission from these sources. Alternatively, in a 3D simulation, a lower background density may be derived from the TeV radial brightness profiles.




\section{Discussion and Conclusions}
\label{dis}

Based on 2D MHD simulations of shell-type supernova remnants evolving in a turbulent background, the evolution of the magnetic field for these remnants can be obtained.  In the leptonic scenario for the $\gamma$-ray emission, the observed correlation between the X-ray and $\gamma$-ray brightnesses implies that the energy density of high-energy electrons is proportional to that of the magnetic field. One can then model the multi-band nonthermal emission with the MHD simulations with the energy partition between energetic electrons and the magnetic field as a free parameter. It is shown that the radial profile of the TeV emission is very sensitive to the density of the background plasma. With reasonable parameters for four TeV bright shell-type SNRs, we reproduce the azimuthal averaged radial brightness profile of the TeV emission obtained with the HESS observations.

Using multi-dimensional hydrodynamic simulations, \citet{Oea14} interpreted the inhomogeneities of the emission from the SNR RX J0852.0-4622 by assuming the blast wave evolved in a cloudy environment. To reproduce the morphology of two opposite lobes for the SNR Kepler, \citet{TS13}
proposed that jets existed before the explosion, which can produce the axisymmetric structure with lobes.
Our simulations are intrinsically 2D, and a 3D model including the details of both the ejecta and the background medium around the
TeV remnants is needed to reproduce the detailed structure of nonthermal emission
of these four remnants studied here.

\begin{acknowledgements}

We thank Dr. Fan Guo for providing a code to generate turbulent ISM. This work is partially support by the Strategic Priority Research
Program ¨C The Emergence of Cosmological Structures of the Chinese Academy
of Sciences, Grant No. XDB09040200,
NSFC grants: 11163006, 11173064, 11233001, and 11233008.
Fang is supported by the NSFC grant (11103016), and by the Key Project of Chinese Ministry of Education (212160). HL is supported by the LDRD program at LANL. 
\end{acknowledgements}


\begin{thebibliography}{}

\bibitem[Abdo et al. (2011)]{ab11}
Abdo, A. A. et al. 2011, ApJ, 734,28
\bibitem[Acero et al. (2009)]{ac09}
Acero, F., Ballet, J., Decourchelle, A., et al. 2009, A\&A, 505, 157
\bibitem[Ackermann et al. (2013)]{ac13}
Ackermann, M., et al. 2013, Science, 339, 807-811
\bibitem[Aharonian et al. (2005)]{ah05}
Aharonian, F. et al. 2005, A\&A, 437,L7
\bibitem[Aharonian et al. (2006)]{ah06}
Aharonian, F. et al. 2006, A\&A, 449, 223
\bibitem[Aharonian et al. (2007)]{ah07}
Aharonian, F. et al. 2007, ApJ,661,236
\bibitem[Aharonian et al. (2008)]{ah08}
Aharonian, F., Akhperjanian, A. G., Barres de Almeida, U., et al. 2008, A\&A,
477, 353
\bibitem[Aharonian et al. (2009)]{ah09}
Aharonian, F. et al. 2009, ApJ,692,1500
\bibitem[Abramowski et al. (2011)]{ak11}
Abramowski A. et al. 2011, A\&A, 531,1
\bibitem[Bamba et al. (2005)]{byh05}
Bamba,A., Yamazaki,R. \& Junko S. Hiraga 2005, ApJ,632,294
\bibitem[Bamba et al.(2012)]{bm12}
Bamba,A. et al. 2012, ApJ, 756,149
\bibitem[Bell (2004)]{b04}
Bell, A. R. 2004, MNRAS, 353, 550
\bibitem[Berezhko \& V\"{o}lk (2010)]{bv10}
Berezhko,E. \& V\"{o}lk 2010, A\&A, 511,34
\bibitem[Bocchino et al. (2000)]{bet00}
Bocchino, F., Vink, J., Favata, F., Maggio, A., \& Sciortino, S. 2000, A\&A, 360, 671
\bibitem[Burlaga \& Lazarus (2000)]{bl00}
Burlaga, L. F., \& Lazarus, A. J. 2000, J. Geophys. Res., 105, 2357
\bibitem[Bykov et al. (2008)]{bu08}
Bykov, A. M., Uvarov, Y. A., \& Ellison, D. C. 2008, ApJ, 689, L133
\bibitem[Cassam-Chena\"{i} et al. (2004)]{cc04}
Cassam-Chena\"{i}, G., Decourchelle, A., Ballet, J., et al. 2004, A\&A, 427, 199
\bibitem[Ellison et al. (2010)]{e10}
Ellison, D. C., Patnaude, D. J., Slane, P., \& Raymond, J. 2010, ApJ, 712, 287
\bibitem[Ellison et al. (2012)]{e12}
Donald C. Ellison, Patrick Slane, Daniel J. Patnaude, \& Andrei M. Bykov 2012, ApJ, 744, 39
\bibitem[Fan et al. (2010)]{fl10}
Fan, Z. H., Liu, S. M., \& Fryer, C. L. 2010, MNRAS, 406, 1337-1349
\bibitem[Fukui et al. (2003)]{f03}
Fukui, Y., Moriguchi, Y., Tamura, K., et al. 2003, PASJ, 55, L61
\bibitem[Giacalone \& Jokipii (1999) ]{gj99}
Giacalone, J., \& Jokipii, J. R. 1999, ApJ, 520, 204
\bibitem[Giacalone \& Jokipii (2007) ]{gj07}
Giacalone, J., \& Jokipii, J. R. 2007, ApJ, 663, L41
\bibitem[Guo et al. (2012)]{gs12}
Fan Guo, Shengtai Li, Hui Li, Joe Giacalone, J. R. Jokipii, \& David Li 2012,ApJ, 747,98
\bibitem[Inoue et al.(2009)]{i09}
Inoue, T., Yamazaki, R., \& Inutsuka, S. 2009, ApJ, 695, 825
\bibitem[Katz \& Waxman (2008)]{kw08}
Katz, B., \& Waxman, E. 2008, JCAP, 01, 018
\bibitem[Kesteven \& Caswell (1987)]{kc87}
Kesteven, M. J., \& Caswell, J. L. 1987, A\&A, 183, 118
\bibitem[Koyama et al. (1995)]{kyet95}
Koyama, K., et al. 1995, Nature, 378, 225
\bibitem[Lemoine-Goumard et al. (2012)]{l12}
Lemoine-Goumard, M., Renaud, M., Vink, J., Allen, G. E., Bamba, A., Giordano, F., \& Uchiyama, Y. 2012, A\&A, 545, A28
\bibitem[Li et al. (2011)]{ll11}
Li, H., Liu, S., \& Chen, Y. 2011, ApJ, 742,L10
\bibitem[Liu et al. (2008)]{lf08}
Liu, S., Fan, Z., Fryer, C. L., Wang, J. M., \& Li, H. 2008, ApJ, 683, L163
\bibitem[Mignone et al. (2007)]{m07}
Mignone, A., Bodo, G., Massaglia, S. et al. 2007, ApJS, 170, 228
\bibitem[Moriguchi et al. (2005)]{m05}
Moriguchi, Y., Tamura, K.,\& Tawara, Y., et al. 2005, ApJ, 631, 947
\bibitem[Morlino \& Caprioli (2012)]{mc12}
Morlino, G. \& Caprioli, D. 2012, A\&A, 538,81
\bibitem[Morlino et al. (2009)]{mab09}
Morlino, G., Amato, E.\& Blasi, P. 2009, MNRAS, 392,240

\bibitem[Obergaulinger et al.(2014)]{Oea14}
Obergaulinger, M., Iyudin, A. F., M¨¹ller, E., Smoot, G. F. 2014, MNRAS, 437, 976

\bibitem[Pacholczyk (1970)]{p70}
Pacholczyk, A. G. 1970, Radio Astrophysics (San Francisco: Freeman)
\bibitem[Pfeffermann \& Aschenbach (1996)]{pa96}
Pfeffermann, E. \& Aschenbach, B. 1996, in Roentgenstrahlung from the
Universe, ed. H. U. Zimmermann, J. Tr¨¹mper, \& H. Yorke, 267
\bibitem[Pisarskietal et al. (1984)]{p84}
Pisarski, P. L., Helfand, D. J., \& Kahn, S. M. 1984, ApJ, 277, 710
\bibitem[Porter et al. (2006)]{p06}
Porter, T. A., Moskalenko, I. V., \& Strong, A. W. 2006, ApJ, 648, L29-L32
\bibitem[Riquelme \& Spitkovsky (2009)]{rs09}
Riquelme, M. A., \& Spitkovsky, A. 2009, ApJ, 694, 626
\bibitem[Rosado, et al. (1996)]{r96}
Rosado, M., Ambrocio-Cruz, P., Le Coarer, E., \& Marcelin, M. 1996, A\&A,
315, 243
\bibitem[Sano et al.(2010)]{s10}
Sano, H., Sato, J., Horachi, H., et al. 2010, ApJ, 724, 59
\bibitem[Smith (1997)]{s97}
Smith, R. C. 1997, AJ, 114, 2664
\bibitem[Tanaka et al (2008)]{t08}
Tanaka, T., Uchiyama, Y., Aharonian, F. A., et al. 2008, ApJ, 685, 988
\bibitem[Tian, et al. (2008)]{tian08}
Tian, W. W., Leahy, D. A., Haverkorn, M., \& Jiang, B. 2008, ApJ, 679, L85
\bibitem[Tsebrenko \& Soker(2013)]{TS13}
Tsebrenko, D., \& Soker, N. 2013, MNRAS, 435, 320

\bibitem[Uchiyama,et al.(2007)]{u07}
Uchiyama, Y., Aharonian, F. A., Tanaka, T., Takahashi, T., \& Maeda, Y. 2007, Nature, 449, 576
\bibitem[Williams et al. (2011)]{w11} Williams, B. J. et al. 2011, ApJ, 741, 96(15pp)
\bibitem[Vink et al. (2006)]{v06}
Vink, J., Bleeker, J., Van Der Heyden, K., Bykov, A., Bamba, A., \& Yamazaki,
R. 2006, ApJ, 648, L33
\bibitem[Yamazaki et al. (2009)]{ykk09}
Yamazaki, R., Kohri, K.\& Katagiri, H. 2009, A\&A, 495,9
\bibitem[Yang \& Liu (2013)]{yl13}
Yang,Chuyuan \& Liu, S. M. 2013, ApJ, 773,138
\bibitem[Yang et al. (2014)]{yz14}
Yang,R. Z., Zhang, X., Yuan, Q., \& Liu, S. M. 2014, A\&A, 567, A23 (7pp)
\bibitem[Yuan et al. (2011)]{yl11}
Yuan, Q., Liu, S. M., Fan, Z. H., Bi, X. J., \& Fryer, C. L. 2011, ApJ, 735, 120(9pp)
\bibitem[Yuan et al. (2012)]{ylb12}
Yuan, Q., Liu, S. M., \& Bi, X. J. 2012, ApJ, 761, 133
\bibitem[Yuan et al. (2014)]{yh14}
Yuan, Q., Huang, X. Y., Liu, S. M., \& Zhang, B. 2014, ApJL, 785, L22(5pp)







\end{thebibliography}
\end{document}